\documentclass[aps,pre,twocolumn,groupedaddress]{revtex4}

\usepackage[dvips]{graphicx,color}

\begin{document}


\title{Revolving rivers in sandpiles: from continuous to intermittent
flows}











\author{E. Altshuler$^{1,2}$, R. Toussaint $^{2}$, E. Mart{\'i}nez$^{1}$, O.
 Sotolongo-Costa$^{1}$, J. Schmittbuhl $^{2}$ and K. J. M{\aa}l{\o}y$^{3}$}

\affiliation{
$^1$``Henri Poincar{\'e}" Group of Complex Systems,
Physics Faculty, University of Havana, 10400 Havana, Cuba.\\
$^2$ Institute of Globe Physics in Strasbourg (IPGS), UMR 7516
CNRS, Universit{\'e} Louis Pasteur, 5 rue Descartes, F-67084 Strasbourg Cedex, France. \\$^3$
Department of Physics, University of Oslo, P.O.Box 1048 Blindern,
 0316
Oslo, Norway.}



\date{\today}

\begin{abstract}

In a previous paper [Phys. Rev. Lett. 91, 014501 (2003)], the
mechanism of ``revolving rivers" for sandpile formation is
reported: as a steady stream of dry sand is poured onto a
horizontal surface, a pile forms which has a river of sand on one
side flowing from the apex of the pile to the edge of the base.
For small piles the river is steady, or continuous. For larger
piles, it becomes intermittent. In this paper we establish
experimentally the ``dynamical phase diagram" of the continuous
and intermittent regimes, and give further details of the piles
``topography", improving the previous kinematic model to describe
it and shedding further light on the mechanisms of river
formation. Based on experiments in Hele-Shaw cells, we also
propose that a simple dimensionality reduction argument can
explain the transition between the continuous and intermittent
dynamics.


\end{abstract}



\maketitle


\section{Introduction}

\begin{figure}[b]
\includegraphics[height=3.0in, width=3.4in]{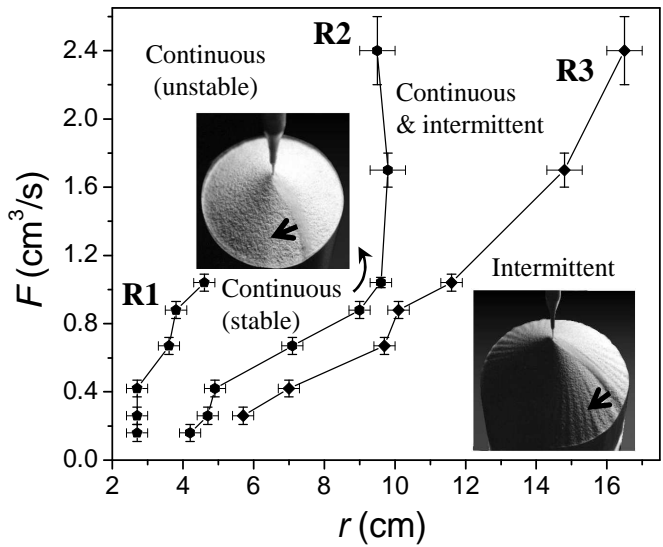}
\caption{\label{fig:f1} Occurrence of continuous and intermittent
rivers, as function of the pile radius $r$ and the input flux $F$.
The upper and lower insets show snapshots of the pile
during the (stable) continuous and the intermittent regimes,
respectively. The arrows indicate the direction of revolution
around the pile}
\end{figure}

When most granular materials are poured onto a horizontal surface,
a conical pile builds up through an avalanche mechanism involving
all the surface of the pile that ``tunes" the angle of repose
about a certain critical value. In a previous paper, however, we
reported a different mechanism of sandpile formation
\cite{Altshuler2003}. In those experiments, as the pile grows, a
river of sand spontaneously builds up, flowing down one side of
the pile from the apex to the base. The river then begins to
revolve around the pile, depositing an helical layer of material
with each revolution, causing the pile to grow. Below a certain
pile size and input flux, the rivers are continuous, and the
surface of the pile is smooth (see upper inset in Fig. 1). The
revolving direction can be either clockwise or counterclockwise,
only depending on random fluctuations in the first stage of the
experiment. For bigger piles, the rivers are intermittent, and the
surface of the pile become undulated (see lower inset in Fig. 1).
X.-Z. Kong and coworkers \cite{Kong2006} reported a detailed
computational model for the revolving rivers which manages to
reproduce many of their features in the continuous regime, but the
intermittent regime was not accounted for. They also noticed that
the details of the grain-grain interactions must be tuned in order
to get the revolving rivers, which matches the experimental fact
that not all sands display such behavior. In this paper, we
explore experimentally the ``dynamic phase diagram" of the
different regimes in our sandpiles (i.e., the regions of
experimental parameters where they appear). We report the details
of the ``topography" of the sandpiles in the continuous and
intermittent regimes, and present a refined ``kinetic" description
of the observed phenomena. We then propose an explanation for the
existence of two types of revolving rivers based on an analogy
with measurements performed in a simpler geometry (i.e., the
Hele-Shaw cell).


\section{The ``dynamic phase diagram" of the revolving rivers}

We used sand with a high content of silicon oxide from Santa
Teresa (Pinar del R\'{\i}o, Cuba) \cite{Altshuler2003,Etien2007}.
The grain size distribution is basically a Gaussian distribution
centered at $200 \mu$m, with a half-width slightly smaller than
$200 \mu$m. From now on, we will adopt the labelling of reference
\cite{Altshuler2003} for the boundary conditions used in each
experiment: BCI stands for a pile grown on a closed cylindrical
container with a flat, horizontal bottom (like a glass of rum),
and BCII corresponds to a pile formed on an open, flat horizontal
surface. To obtain the dynamic phase diagram, we grew piles in
boundary condition BCII. We used glass funnels of various
diameters to obtain different values of the input flux. It was
determined in each experiment by measuring the volume of the
conical pile, and dividing it by the time elapsed since the first
grains were dropped on the table. For a given input flux, three
different pile base radii where identified by simple observation
as the pile grew. Below $r1$, continuous rivers appeared and
disappeared (as eventually happens in common sands), so we will
refer to them as \textit{unstable}. Above $r1$ and below $r2$, a
\textit{stable} continuous river revolved around the pile in a
given direction, without any major changes in its shape. After
$r2$, the continuous river could eventually become intermittent:
the downhill flow would suddenly stop at the edge of the pile, and
a ``finger" of sand would escalate from bottom to top, like a
``stop-up" front \cite{Douady2000}. All in all, the continuous or
intermittent rivers would keep revolving around the pile with the
frequency described in \cite{Altshuler2003} (see also Fig. 7).
However, at some point this process transformed again into a
continuous, revolving river. As the radius of the pile grew from
$r2$ to $r3$, the intermittent mechanism took place within larger
and larger time intervals, while the continuous mechanism slowly
disappeared. At $r3$, intermittent rivers where established as the
only dynamical phase in the pile. As $r1$, $r2$ and $r3$ were
determined for various input fluxes, three boundaries $R1$, $R2$
and $R3$ were established to separate the different dynamical
phases, as shown in Fig. 1. We have observed that the positions of
these lines are influenced by the distance between the funnel that
delivers the sand and the tip of the pile, but we kept this
distance fixed at approximately $2$ cm during the experiments. It
should be noticed that, for input fluxes bigger than approximately
$1 cm^3/s$, stable continuum rivers are never established, and a
direct transition from unstable continuous to intermittent rivers
takes place through $R2$.

\begin{figure}[b]
\includegraphics[height=5in, width=3.3in]{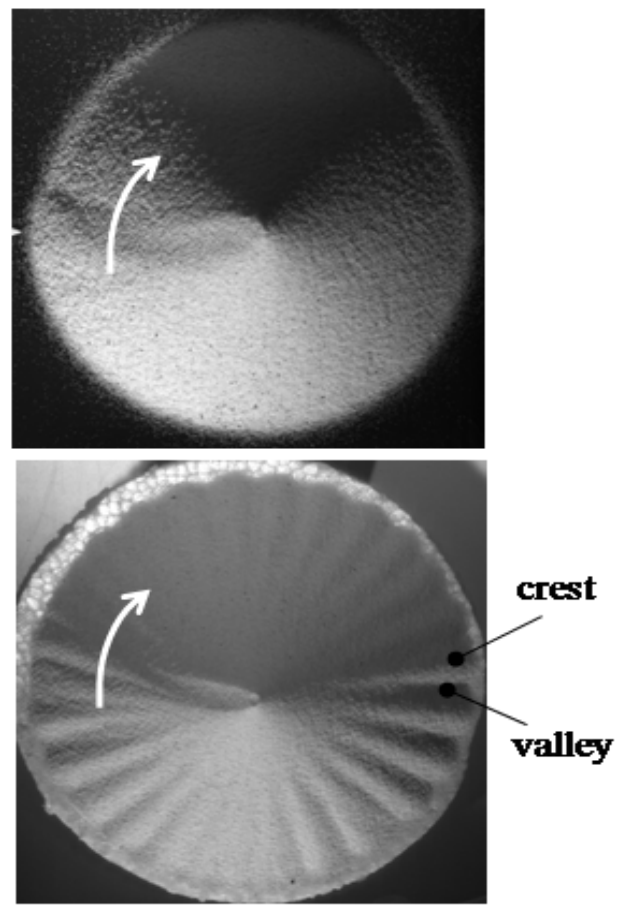}
\caption{\label{fig:f2} Topography of the continuous and
intermittent regimes. Upper panel: top view showing a pile
``frozen" during a continuous river experiment. The diameter of
the pile is $11$cm. Lower panel: top view of a pile ``frozen"
during an intermittent river experiment. The diameter of the pile
is $17$cm --surface undulations extend approximately $6$ cm from
the edge of the pile in the radial direction. Black arrows on top
of the flowing river indicate the direction of revolution of the
rivers as sand is added from above.}
\end{figure}

\section{Refined kinematic model}

Figure 2 shows images taken from top of ``frozen" piles in the
stable continuous (upper panel) and intermittent (lower panel)
regimes obtained for boundary condition BCII. Piles were formed
using a steady input flux, which was just switched off to take the
pictures. In both images, the river can be identified as a curved
groove from the apex to the base of the pile, in a general
vertical direction, below the apex. In order to quantify the
topography of the piles (at least for one size), we measured its
slope angle (polar angle) versus the angle around the pile
(azimuthal angle). The measurement was performed by carefully
setting a ruler tangential to the slope of the pile at different
azimuthal angles, taking a lateral picture of the system, and
calculating the angles between the ruler and the horizontal from
the resulting images. This ruler was set along the section of the
slope where the angle is essentially constant (this angle becomes
smaller at a few cm from the top of the piles). Figure 3 shows how
the slope angle depends on the azimuthal angle for the case of the
continuous rivers. Contrary to the first impression when the
picture of the upper panel of Fig. 2 is examined, there is a wide
angular zone of variation of the slope angle, quantified in Fig.
3. This suggests that, while most of the downhill flow is
concentrated into the relatively narrow river clearly visible
below the apex of the pile (upper panel of Fig. 2), there is a
wider area of grains rolling downhill. The slope angle is
typically around 36.5 degrees far from the river. It significantly
decreases to 33 degrees within the river. The width of the low
slope angle region extends over $150$ degrees along the azimuthal
direction. This corresponds to a length of around $28$ cm along
the circumference of the pile, which is much larger than the river
(which is a few cm wide). To illustrate this, Figure 4 shows the
difference between two images of a sandpile grown in BCI in the
continuous regime, separated by a lapse of $0.15$ seconds (the
video was taken by a \emph{Photon Fastcam Ultima-ADX model 120K},
and the image was obtained using the ``difference" option of {\it
ImageJ}). Besides the fast flow along the main stream of the
river, whiter and darker spots indicate the downhill movement of
sand at a smaller speed behind the main stream area.

 The
situation is quite different in the case of the intermittent
regime. The measurement of the angles of repose of several
``valleys" and ``crests" corresponding to the picture in the lower
panel of Fig. 2 does not indicate a wide transition in the angle
of repose, as in the case of the continuous rivers, but suggests
two basic polar angles: $37.1 \pm 0.16$ degrees, for the valleys,
and $36.1 \pm 0.20$ degrees for the crests. It should be noticed,
however, that the pile in the intermittent regime contains an
inner circle where the surface is smooth, and an outer ring of
undulating surface, as seen in Fig. 2 (lower panel). The values we
have reported here for the slope angles at the valleys and crests
correspond to this ring of approximately $6$ cm width in the
radial direction, while the average slope of the pile in the inner
circle is harder to measure due to rounding near the apex of the
pile.

\begin{figure}
\includegraphics[height=2.5in, width=3.3in]{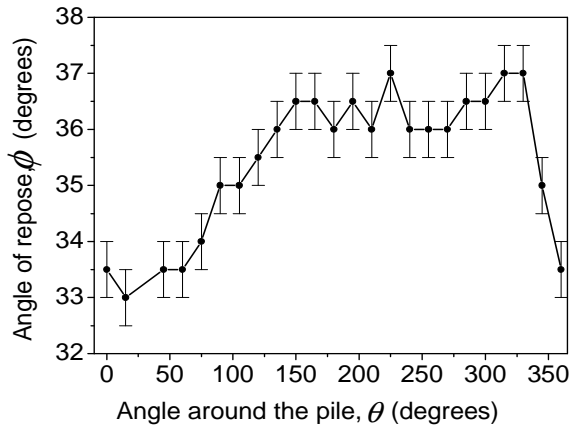}
\caption{\label{fig:f3} Dependence of the angle of repose on the
azimuthal angle for the picture of the upper panel of Fig. 2
(continuous river). The ``0" of the azimuthal angle has been taken
along the vertical diameter under the apex, and it increases
clockwise. The error bars correspond to the largest variation of
the measured angles of repose in three different attempts on the
same pile.}
\end{figure}

\begin{figure}
\includegraphics[height=2.5in, width=3.3in]{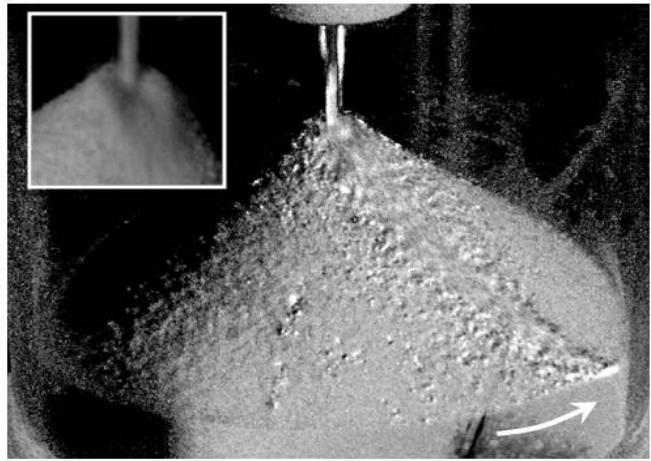}
\caption{\label{fig:f4} Moving grains in the continuous river
regime. The figure is the difference between two pictures of a
continuous river separated by $0.15$ seconds (the river was
revolving around the pile at approximately $0.1$ turns/s). This
difference is superimposed with the first of these two pictures.
The arrow indicates the direction of the revolution of the river
around the pile. The inset shows a raw picture of the crater, at
the upper end of the pile.}
\end{figure}

\begin{figure}
\includegraphics[height=4.5in, width=3in]{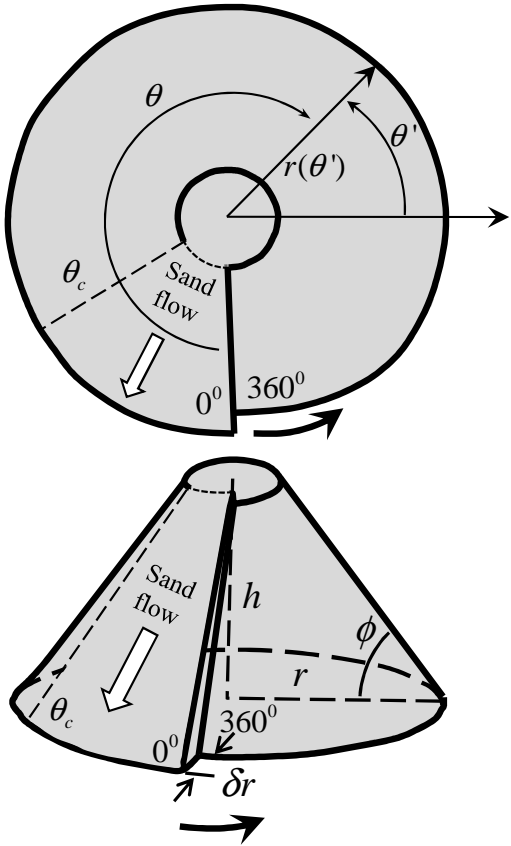}
\caption{\label{fig:f5} Diagram of the model used to estimate the
revolving speed of the rivers, and the polar bottom angle versus
the azimuthal angle. Upper panel: top view of the pile model.
Lower panel: spatial view of the pile model (notice that the angle
of the flowing zone has been decreased in this representation
relative to the top view, to improve drawing clarity). The black
arrows indicate the direction of revolution around the pile.}
\end{figure}

The phenomenological model for the revolving rivers presented in
reference \cite{Altshuler2003} is able to predict the radius and
time dependence of the frequency of revolution of the piles, but
oversimplifies the geometry of the sandpile in such a way that the
angle of repose is constant around the pile, in contradiction with
the real topography shown in Fig. 2 and quantified in Fig. 3. We
present here a more realistic model, which is schematized in
Fig.~\ref{fig:f5} for BCII. This model has a number of advantages
relative to the one presented in reference \cite{Altshuler2003}.
First, it takes into account that the apex of the pile is not
sharp, but presents a small ``crater" (very exaggerated in the
schematics shown in Fig. 5). Second, the river, located in the
section on the cone on which the sand flows, has a nonzero width.
Third, the angle of repose is not constant around the pile, but
can be shaped as depicted in Fig. 5, where the width of the
downhill flowing area has been tuned to $150$ degrees in the
azimuthal direction (the figure displays a smaller angle). We add
two further constraints, which roughly match experimental
observations: the characteristic length $\delta$r is constant, and
so is the radius of the crater.

The geometry of the pile, in the continuous regime,
can be schematized as follows:

We define an azimuthal coordinate $\theta$ around the pile in a
frame co-moving
 with the revolving river,
so that $\theta=0$ corresponds to the position
of the flowing river, where most of the flow is located.
In the reference frame of the laboratory, the azimuthal coordinate
 $\theta'$ is
$\theta'= \theta_0(t) - \theta$, where  $\theta_0(t) = \theta_0(0) +
 \int_0^t \omega (t') dt'$ is the azimuthal angle of the river,
and $ \omega(t)=\dot{\theta}_0(t)$ is the instantaneous angular
 velocity of the revolving river.

The radius of the bottom of the pile is fixed after the river has added
 a layer of thickness $\delta r$: this radius is denoted
$r(\theta')$. There is no evolution of this radius apart from the
 outlet of the concentrated river, i.e. any other surface flow than the river
 on the sides of the pile does not reach the bottom.
This radius jumps by a characteristic length $\delta
 r=r(\theta'+360^{\circ})-r(\theta')$ at each passing of the river,
 which adds a new layer of grains
to the pile. This
 jump corresponds to the distance over which grains flowing down the
 pile at a characteristic speed, dissipate entirely their kinetic energy
 over the flat bottom. This jump has a characteristic value around
 $4$mm.

 Calling h(t) the height of the center
 of the pile,  there is a small crater forming around this tip.
The crater corresponds then to a sharp crest around the center of the
 pile,
 slightly higher than the central point, and exists for azimuthal
 angles $\theta$
 between $\theta_c$ and $360^{\circ}$ . A characteristic value observed
 for $\theta_c$
is around $150^{\circ}$, as pointed out before. There is no
observed surface flow on the sides
 of the pile below this crater crest,
i.e. the surface of the pile corresponding to
 $\theta_c<\theta<360^{\circ}$ is frozen.
The crater crest lies at a height $\delta h(\theta)$ above the center,
 i.e. at an altitude $h(t)+\delta h(\theta)$:
the crater shape revolves at the same speed as the flowing river.
In the reference frame of the bottom plate, a point of the crest
is
 also fixed, i.e.
$h(t)+\delta h(\theta_0(t)-\theta')$ is constant, so its time
derivative is $\dot{h} + (d \delta h / d \theta) \omega =0$: the
crater has a screw
 like shape, with an azimuthal slope of the crest
$(d \delta h / d \theta)=-\dot{h}/\omega$ set by the ratio between the
 pile rising speed and the revolving angular velocity.

After the passing of the river, there is a small surface flux, visible
 as a few whiter and darker dots on Fig.~\ref{fig:f4}, increasing the effective angle of
 the slope, as seen in Fig.~\ref{fig:f3}.
 This surface flow happens for azimuthal angles between
$0^{\circ}$ (the river) and $\theta_c$. The tip of the pile, for
these angles, presents no change of sign of the radial slope (no
crater crest). For $\theta$ larger than $\theta_c$, the crest of
the crater at the tip of the pile is formed around the central
point,
 i.e. noting $S(r,\theta$) the coordinates of the surface of the pile,
the radial slope of $\partial S/\partial r$ is positive for low
radii $r$, and negative for larger ones,
 after the crest.
At these angles, this crest prevents further surface flow on the
 sides of the pile, so that the rest (polar) angle $\Phi_0(\theta)$ is
 frozen at a roughly constant value between $\theta_c$, (the beginning
 of the crest) and $\theta=360^{\circ}$.
To illustrate this organization, a collection of schematic cuts of the
 pile side at various azimuth $\theta$ is shown in Fig. 6.

\begin{figure}
\includegraphics[height=5.2 in, width=3.4in]{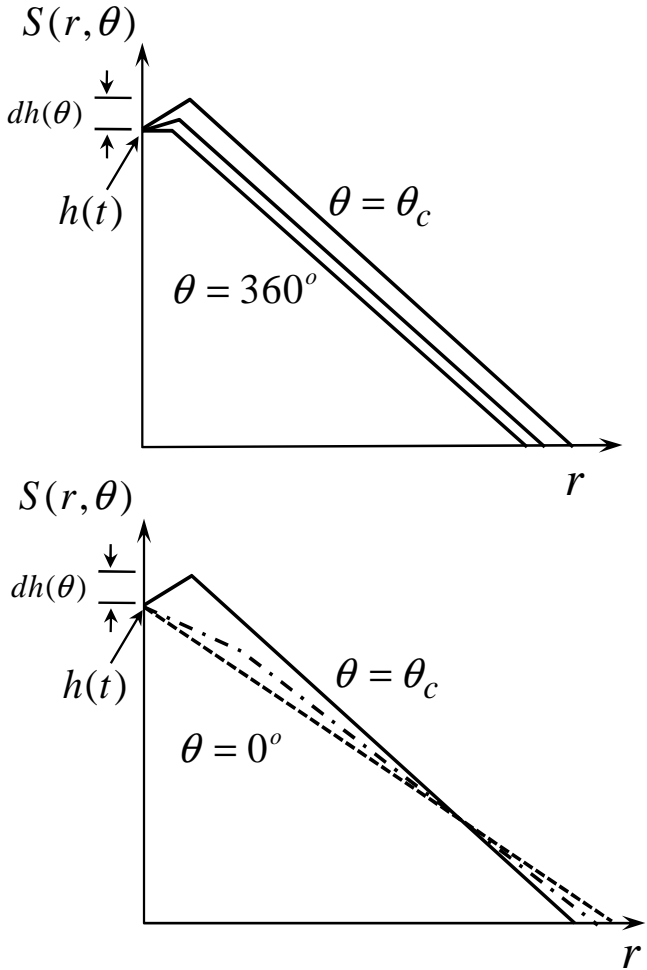}
\caption{\label{fig:sidepile} Cuts of the pile side at constant
azimuth $\theta$ in the continuous regime. Upper panel: frozen
side of the pile, below a crater crest. Lower panel: part where
the revolving river passes, and where the surface flow takes place
behind it.}
\end{figure}

The role of this crater on the organization of the flow in
 revolving rivers seems thus very important. For example, when the incoming
 sand flux presents too many lateral oscillations or is too wide, the
 crater and revolving river formation are prevented. Similarly, when the pile
 has not grown enough to reach a diameter beyond the distance of
 saltation from the impact point, the crater formation is prevented, and
 unstable rivers are observed -i.e., beyond the line R1 in Fig.~\ref{fig:f1}.
The fact that grains do not bounce too far when they arrive on the
 pile, i.e. a moderate restitution coefficient, is certainly an important
 ingredient to allow the crater formation.

\begin{figure}
\includegraphics[height=5.2 in, width=3.0in]{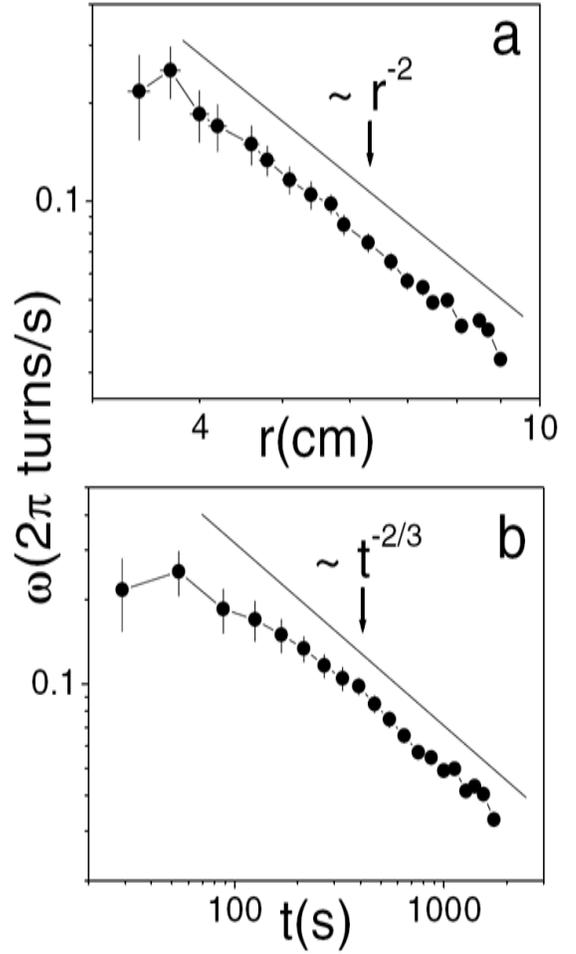}
\caption{\label{omega} Radius (a) and time (b) dependencies of the
frequency of revolution of the rivers (taken from
\cite{Altshuler2003}). The points correspond to experimental
values.}
\end{figure}

From this geometry, one can also infer a relationship between the pile
 radius and the angular velocity, as reported in \cite{Altshuler2003}: during
 a time $dt$, a volume of sand $F dt$ is deposited over a layer of
 size $\delta r \tan(\Phi_0) \omega r^2 dt / 2$, where $\Phi_0$ is the respose angle
along most of the pile (for azimuth $\theta$ between $\theta_c$ and $360^{\circ}$).
Consequently, mass conservation of the grains imposes that $\omega = 2 F / [r^2
 \delta r \tan(\Phi_0)]$. This power-law dependence of the angular velocity
 over the pile radius is verified experimentally, as shown in Fig.~\ref{omega}.
Since the pile radius increases from flux conservation so that
$\pi r(t)^3\tan(\Phi_0)/3=Ft$, i.e. as $r(t)=[3Ft/\pi
\tan(\Phi_0)]^{1/3}$, the angular velocity goes as
\begin{equation}
\omega = (\pi/3)^{2/3}\frac{2 F^{1/3}}{\delta r
(\tan{\Phi_0})^{1/3}} \frac{1}{t^{2/3}}.
\end{equation}
This power law decrease of the angular velocity with time is also
verified experimentally (Fig. 7).

At a certain critical size of the pile radius $r_{ci}$, the flow
becomes
 intermittent and exhibits avalanches, while still being concentrated
 along a river. These avalanches extend over a basis $\delta r$ from the
 pile bottom, comparable to the
horizontal extent of layers added by the flowing river in the
continuous regime.
 These avalanches form delta-shaped piles on the side of the main cone,
 with a slightly carved river penetrating above the apex of the delta
 (as seen below the pile's apex, lower panel of Fig. 2) which is
 ``smoothen out" as the river moves on laterally.
The slope angle of these delta is lower than the average angle
around the
 pile, with typically a stop angle $\Phi_1=33^{\circ}$, as on the
 flowing part of the continuous regimes, whereas the rest angle for the
 average slope of the pile is around $\Phi_0=36
 ^{\circ}$ --see
Fig. 3. The radial extent $l$ of these deltas from the bottom of
the pile
 is set by the geometrical condition $(l+\delta r) \tan(\Phi_1) = l
 \tan(\Phi_0)$ (see Fig. 8), which corresponds to a constant value $l \sim 6 cm$, as
 observed on Fig.~\ref{fig:f2}. Once these deltas are filled to the top, as
 well as the slightly carved zone above, the river feeding this delta
 revolves and a new delta is formed by an avalanche developing next to the
 previous one.

 \begin{figure}
\includegraphics[height=2.4 in, width=3in]{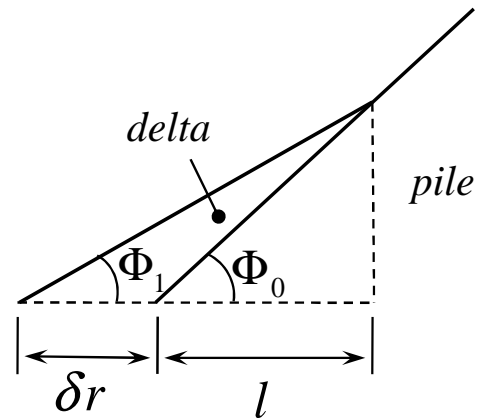}
\caption{Cuts of the pile near the lower boundary in the
intermittent regime.}
\end{figure}

\section{Continuous vs. intermittent regimes: comparison with Hele-Shaw
 experiments}

The transition from continuous to intermittent granular flows have
been extensively studied using two basic geometries: the rotating
drum \cite{Rajchenbach1990} and the Hele-Shaw cell
\cite{Durian2000}. Typically, the intermittent (or ``avalanche")
flow is transformed into continuous flow by increasing the
rotation speed of the drum or the input flux, respectively.  In
the case of the Hele-Shaw cell arrangement, a transition from
continuous to intermittent flow (start down and stop up fronts
\cite{Douady2000}) takes place when the heap reaches a critical
size, even if the input flux and channel width are kept constant.
Here we explore such transition in our sand in order to establish
an analogy with the sandpile scenario. The cell consisted in a
horizontal base and a vertical side-wall, sandwiched between two
square glass plates with inner surfaces separated by a distance
$w$ in the range from $0.4$ cm to $1.2$ cm. The lengths of the
base and the vertical wall were approximately $d\approx 22$ cm.
The sand was poured near the vertical wall using a sliding window
of width $w$ and variable aperture along the horizontal direction.
Although in most of the results we report below the distance
between the funnel and the upper side of the heap varied from
$10$cm to $1$cm during a typical experiment, we performed some
tests keeping the distance fixed at $2$cm, getting similar
results. The input volume per unit time, $F$, was calibrated by
calculating the volume of sand deposited in the cell in a given
time interval.

\begin{figure}
\includegraphics[height=3.2 in, width=2.4 in]{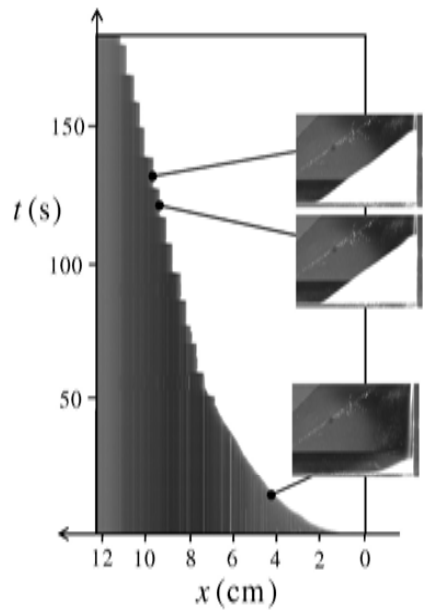}
\caption{\label{fig:f6} Spatio-temporal diagram of the growing
heap. The diagram was made for a sequence of images taken at every
$0.1$s to a heap growing in a Hele-Shaw cell with $w=4$ mm and
$F=0.125$$cm^3/s$. The cuts where taken along an approximately
$12$ cm-long horizontal line near the bottom of the cell. As the
time goes by, the heap grows continuously, corresponding to a
smooth increase of the white area in the diagram. At a critical
distance of $7$ cm, avalanches start, and the diagram shows a
step-like behavior. The insets on the right contains real pictures
of the heap illustrating different moments of the spatio-temporal
diagram.}
\end{figure}

Fig. 9 shows a spatio-temporal diagram of evolution of one heap
formed into a cell with $w=4$ mm using an input flux of $0.25$
$cm^3/s$. It is based on a digital video acquired with  an
\emph{Optronics Camrecord 600}, some of which snapshots are shown
as insets. A $15$cm-long horizontal line of pixels was taken
$0.5$cm from the bottom of the Hele-Shaw cell every $1/10$ of a
second. The spatio-temporal diagram was constructed as a stack of
such lines (the darker zones correspond to the air, while the
clearer ones correspond to the sand). The position of the row of
pixels was taken in such a way, that the interface between dark
and clear zones follows the displacements at the lower end of the
growing heap. As can be seen at low times, the end of the heap
near the bottom of the cell grows smoothly. However, approximately
when the horizontal size of the heap reaches a length of $7$cm
(see arrow in Fig. 9), the spatio-temporal is no longer smooth,
but resembles a ladder: we define in this way the transition
between the continuous and the intermittent (or avalanche)
regimes. In the latter (well described in references
\cite{Douady2000,Mendes1999,Mendes2000}), sudden avalanches roll
downhill (corresponding to the high velocity, almost horizontal
segments in the spatio-temporal diagram), and then a front climbs
the hill as a new layer of sand is added from bottom to top
(corresponding to the zero-velocity, vertical segments in the
spatio-temporal diagram). The general shape of this diagram
resembles Fig. 2 in \cite {Mendes1999}, where the authors use two
equations for thick flows proposed in \cite{DeGennes1998} to model
the heap formation in Hele-Shaw geometry. However, they do not
report a continuous-intermittent transition --only the
intermittent regime is described. We observe, though, other
predictions reported in \cite{Mendes1999}, such as the
``segmented" profiles of the heaps (see the two upper insets in
our Fig. 9) and stratification lines parallel to the free surface
of the heaps. However, stratification is only visible in our case
after the heap has reached the intermittent regime, i.e., where
the model proposed in \cite{Mendes1999} fully applies.



By quantifying the critical length, $x_{ci}$, at which the
transition between continuous and intermittent flow happens, for
different input fluxes, we construct the ``phase diagram"
presented in Fig. 10. There, the vertical axis represents the
input flux as the volume per unit time of sand added to the cell.
As the width of the cell is increased, the transition line
increases its slope, indicating that, for a given $F$, the
intermittent flow is reached for smaller piles.
The inset of Figure 10 shows a superposition of the transition
line for the case of a Hele-Shaw cell of $w = 12 mm$ and the two
lines $R2$ and $R3$ corresponding to the {\it wide} transition to
the intermittent regime in the sandpile geometry. The relative
position of the lines suggests that the revolving rivers can be
taken, to a first approximation, as a $12$-$mm$-wide ``revolving
Hele-Shaw" cell that performs a revolution around the pile a the
same frequency of the rivers \cite{Altshuler2003}. Within this
picture, the role of the river is just to reduce the
dimensionality of the flow in the pile.

\begin{figure}
\includegraphics[height=4 in,
 width=3.1in]{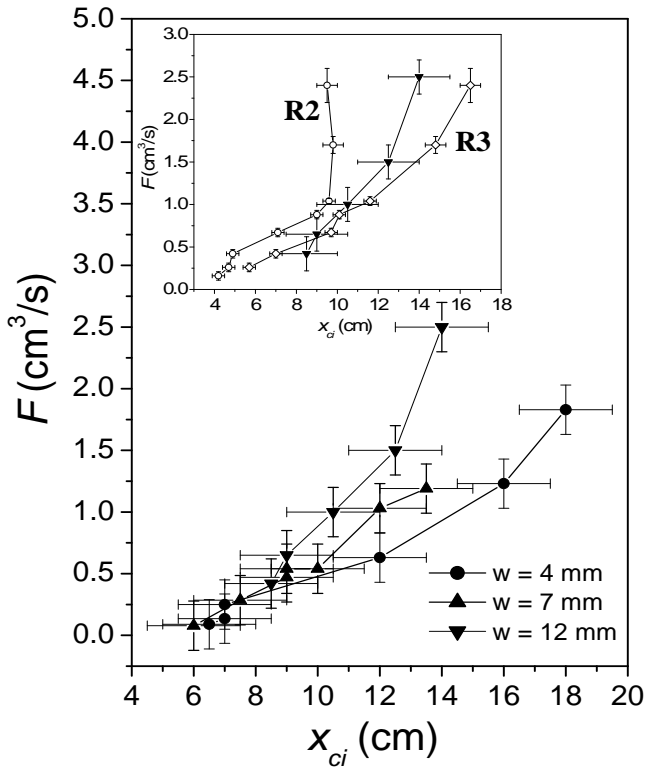}
\caption{\label{fig:f9} Transition from the continuous to the
intermittent regimes in Hele Shaw and revolving rivers
configurations. The vertical axis corresponds to input volume of
sand per unit time, and the horizontal axis is the length of the
base of the heap at the transition. The inset shows a comparison
between the line of a $12mm$ width cell and the lines $R1$ and
$R2$ taken from Fig. 1.} \vspace{-0.5cm}
\end{figure}


\begin{figure}
\includegraphics[height=3.0in, width=3.3in]{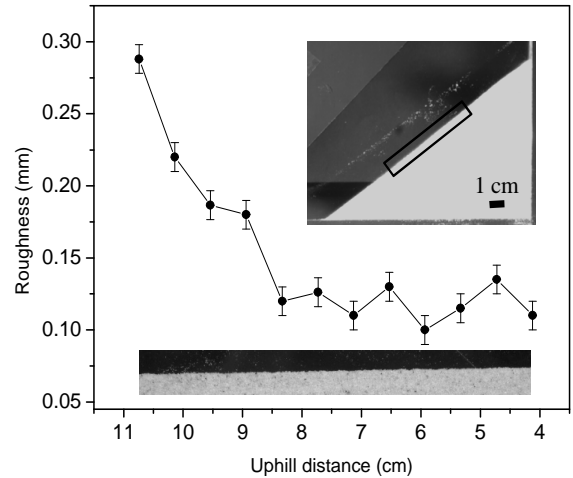}
\caption{\label{fig:f11} Spontaneous segregation. The graph
indicates the decrease in the roughness amplitude of the pile as
we move uphill, suggesting segregation of bigger particles near
the lower edge of the heap. The roughness amplitude has been
measured as the root mean square of the elevation across $12$mm
long running windows on the free surface seen in the $7$cm-long
picture of the lower inset, corresponding to the boxed region
indicated in the upper inset. We define the uphill distance as the
distance from a point on the surface to the top of the pile.}
\end{figure}

The specific mechanism that triggers the transition from the
continuous to intermittent regimes as the size of the system grows
is still unclear for our sandpiles, and also probably for the
Hele-Shaw configuration (we underline that a qualitative
explanation is given in \cite{Douady2000} for a heap of fixed
length). Here we propose an explanation based on segregation. As
well known in the case of debris flows in mountains, bigger rocks
tend to accumulate in the flow front as it slides down. This is
clearly illustrated, for example, in Fig.1.4 of reference
\cite{Taka1991}.
Careful inspection of the free surface of our heaps in the
Hele-Shaw configuration also hints at segregation of big grains as
we move down the slope, even when our grains are not significantly
bi-disperse. In fact, the roughness amplitude of the surface
varies from approximately $0.1$mm to $0.3$mm as we move down the
hill (see Fig. 11). If one assumes that a slope of bigger grains
provides the necessary effective friction to start the growth of a
``stop up" front, we speculate that, when the heap grows to a size
large enough for the segregation of big grains near the base
reaches a certain threshold, the intermittent regime is triggered.
In analogy, the increase in the radius of the pile is a necessary
condition to reach the intermittent regime in the revolving
rivers.

\section{Conclusions}

We have established experimentally the ``dynamical phase diagram"
of the continuous and intermittent regimes for revolving rivers in
sandpiles. One somewhat unexpected feature of the diagram is that
{\it stable} continuous rivers can only exist below a certain
input flux threshold: the intermittent regime is the most robust
dynamics in the system. The details of the pile shape and of the
movement of grains on its surface indicate that, while most of the
downhill flow in the continuous regime takes place within the
``river itself", there is a wide area behind it that contributes
with much smaller flow. This fact, however, plays a relevant role
in the slow change in the slope as one moves around the pile. We
have also improved the kinematic model presented in
\cite{Altshuler2003} in order to mimic in detail the measured
topography of the piles in the continuous regime. The model also
allows to estimate the extension of the undulating pattern in the
intermittent regime based on other experimental parameters. By
performing a series of experiments with Hele-Shaw cells, we
conclude that, due to the reduced dimensionality of the granular
flow along the rivers, we can describe them as ``revolving"
Hele-Shaw cells of $12$ mm width --the analogy accounts both for
the continuous and the intermittent regimes. Finally, we propose
that the segregation of big grains after the pile has reached a
critical size is responsible for the appearance of ``stop-up"
fronts and, consequently, for the transition to the intermittent
regime.

\begin{acknowledgments}

We thank H. Herrmann for inspiration in the exploration of new
sandpile models, O. Ramos, A. Stayer and K. Robbie for
experimental support, and E. Cl\'{e}ment and S. Douady for advice
and useful discussions. We acknowledge support from the French
Norwegian PICS project, the INSU DYETI program, the ANR CTT
ECOUPREF program, the REALISE program, the Alsace region, and the
Cuban PNCIT project 023/04.

\end{acknowledgments}



\end{document}